\documentclass[aps,prc,twocolumn,superscriptaddress,showkeys,nofootinbib]{revtex4-1}  

\usepackage[utf8]{inputenc}

\usepackage{graphicx,color}
\usepackage{amsmath,amssymb,amsfonts}
\usepackage{hyperref}

\usepackage{xspace}	

\def \sc23  {\mbox{$\mathrm{SC}(2,3)$   }\xspace}
\def \sc24  {\mbox{$\mathrm{SC}(2,4)$   }\xspace}

\def \nsc23 {\mbox{$\mathrm{NSC}(2,3)$}\xspace}
\def \nsc24 {\mbox{$\mathrm{NSC}(2,4)$}\xspace}

\begin{document}
\title{Beam Energy Dependence of the Linear and Mode-Coupled Flow Harmonics Using the A Multi-Phase Transport Model}
\medskip
\author{Niseem~Magdy} 
\email{niseemm@gmail.com}
\affiliation{Department of Chemistry, State University of New York, Stony Brook, New York 11794, USA}
\begin{abstract}
In the framework of the A Multi-Phase Transport (AMPT) model, the multi-particle azimuthal cumulant method is used to calculate the linear and mode-coupled contributions to the quadrangular flow harmonic ($\textit{v}_{4}$), mode-coupled response coefficient as a function of centrality in Au+Au collisions at $\sqrt{\textit{s}_{NN}}$ = 200, 39, 27 and 19.6~GeV. 
This study indicated that the linear and mode-coupled contributions to the $\textit{v}_{4}$ are sensitive to the beam energy change. Nevertheless, the correlations between different order flow symmetry planes and the mode-coupled response coefficients show a  weak beam energy dependence.
In addition, the presented results suggest that the experimental measurements that span a broad range of beam energies can be an additional constraint for the theoretical models’ calculations.
\end{abstract}
\maketitle
\section{Introduction}
Experimental investigations of heavy-ion collisions demonstrate the formation of Quantum Chromodynamics (QCD) matter called  Quark-Gluon Plasma (QGP) at the Relativistic Heavy Ion Collider (RHIC) and the Large Hadron Collider (LHC)~\cite{Shuryak:1978ij,Shuryak:1980tp,Muller:2012zq}. One of the primary purposes of previous and present experimental studies of heavy ion collisions is to understand the QGP transport properties, such as the shear viscosity divided by the entropy density $\eta / \textit{s}$~\cite{Shuryak:2003xe,Romatschke:2007mq,Luzum:2008cw,Bozek:2009dw,Acharya:2019vdf,Acharya:2020taj,Adam:2020ymj}.

In the past years, experimental measurements of anisotropic flow resumed being a beneficial route to the extraction of $\eta / \textit{s}$~\cite{Heinz:2001xi,Hirano:2005xf,Huovinen:2001cy,Hirano:2002ds,Romatschke:2007mq,Luzum:2011mm,Song:2010mg,Qian:2016fpi,Magdy:2017ohf,Magdy:2017kji,Schenke:2011tv,Teaney:2012ke,Gardim:2012yp,Lacey:2013eia}. In addition, anisotropic flow gives the viscous hydrodynamic response to the initial-state energy density anisotropy described by the complex eccentricity vectors $\mathcal{E}_{n}$~\cite{Alver:2010dn,Petersen:2010cw,Lacey:2010hw,Teaney:2010vd,Qiu:2011iv}:
\begin{eqnarray}
\mathcal{E}_{n} & \equiv &  \varepsilon_{n} e^{i {\textit{n}} \Phi_{n} }  \\  \nonumber
&\equiv  &
  - \frac{\int dx\,dy\,\textit{r}^{n}\,e^{i {\textit{n}} \varphi}\, \textit{E}(r,\varphi)}
           {\int dx\,dy\,\textit{r}^{n}\,\textit{E}(r,\varphi)}, ~(\textit{n} ~>~ 1), \\
           x &=& r~\cos\varphi, \\
           y &=& r~\sin\varphi, 
\label{epsdef1}
\end{eqnarray}
where  $\varepsilon_{n}$ and ${\Phi_{n}}$ are the magnitude and the angle of the  n$^{th}$ order eccentricity vector, $\varphi$ is the spatial azimuthal angle, and ${\textit{E}}(r,\varphi)$ is the initial anisotropic energy density profile~\cite{Teaney:2010vd,Bhalerao:2014xra,Yan:2015jma}.

The anisotropic flow can be given by the Fourier expansion of the azimuthal anisotropy of particles emitted relative to the collision symmetry planes~\cite{Poskanzer:1998yz}:
 \begin{equation}
\label{eq:1-1}
E \frac{d^{3}N}{d^3p} =  \dfrac{1}{2\pi} \dfrac{d^2N}{p_{T} dp_{T} dy} 
                                             \left( 1 +  \sum^{\infty}_{i=1} 2 \textit{v}_{n}  \cos\left(    n \left( \phi - \psi_{m} \right)    \right)        \right),
\end{equation} 
where $\textit{v}_{n}$ stands for the $n^{th}$ flow coefficient, $y$ is the rapidity,  $\phi$ represents the particle azimuthal angle,   $p_{T}$ gives the transverse momentum, and $\psi_{m}$ is the $m^{th}$ order symmetry plane. The flow harmonics $\textit{v}_{n}$(n=1, 2, and 3) are called directed, elliptic, and triangular flow, respectively.

The prior and current studies of $v_2$ and $v_3$ suggest that, to a reasonable degree, they are linearly related to the medium response~\cite{STAR:2018fpo,ALICE:2016kpq,Adamczyk:2017hdl,Qiu:2011iv, Adare:2011tg, Aad:2014fla, Aad:2015lwa,Magdy:2018itt,Alver:2008zza,Alver:2010rt, Ollitrault:2009ie};
\begin{eqnarray}\label{eq:1-2}
v_{n} = \kappa_{n} \varepsilon_{n},
\end{eqnarray}
where $\kappa_{n}$ (for $n$= 2 and 3) encodes the importance of the QGP's $\eta/s$~\cite{Adam:2019woz,Heinz:2013th}. 
The higher-order flow harmonic $\textit{v}_{4}$~\cite{Magdy:2019ojv,Adam:2019woz,Magdy:2018itt,Adamczyk:2017ird,Magdy:2017kji,Adamczyk:2017hdl,Alver:2010gr, Chatrchyan:2013kba}, will exhibit a linear response to the same-order eccentricity as well as a mode-coupled response to the lower-order eccentricity~$\varepsilon_{{{2}}}$ ~\cite{Teaney:2012ke,Bhalerao:2014xra,Yan:2015jma,Gardim:2011xv}:
\begin{eqnarray}
V_{4}  &=&  v_{4} e^{i 4 \psi_{4}} = \kappa_{4} \varepsilon_{4} e^{4i\Phi_{4}} + \kappa_{4}^{'} \varepsilon^{2}_{2} e^{4i\Phi_{2}}
\nonumber \\
           &=&  V_{4}^{\rm Linear} +   \chi_{4,22} V_{4}^{\rm MC}, \label{eq:1-3a}
\end{eqnarray}
where $\kappa_{4}^{'}$ represents the mixed effect of the medium properties and the coupling between the lower- and higher-order eccentricity harmonics. The $\textit{V}^{\rm Linear}_{4}$, $\textit{V}^{\rm MC}_{4}$, and $\chi_{4,22}$ are the linear and the mode-coupled contributions to $\textit{V}_{4}$ and the mode-coupled response coefficients, respectively.

The mode-coupled response to the $\textit{V}_{4}$ represents additional constraints for the initial-stage dynamics and the $\eta/s$ extraction~\cite{Bilandzic:2013kga,Bhalerao:2014xra,Aad:2015lwa,ALICE:2016kpq,STAR:2018fpo,Zhou:2016eiz,Qiu:2012uy,Teaney:2013dta,Niemi:2015qia,Zhou:2015eya}.
Therefore, ongoing work suggests leveraging comprehensive measurements of $\textit{v}_{4}^{\rm Linear}$ and $\textit{v}_{4}^{\rm MC}$ will provide additional constraints to differentiate between various initial-state models~\cite{Acharya:2017zfg,ALICE:2020sup,CMS:2019nct,Adam:2020ymj,STAR:2018fpo}. In addition, these measurements will pin down the $\eta / \textit{s}$ dependence on temperature ($T$) and baryon chemical potential ($\mu_{B}$).

The paper is organized as follows: section~\ref{Sec:2} describes the AMPT model and the analysis technique employed in this work. Sec.~\ref{Sec:3} conveys the results of this work. The summary is presented in Sec.~\ref{Sec:4}.

\section{Method}\label{Sec:2}
The AMPT~\cite{Lin:2004en} model (version ampt-v2.26t9b) simulated events are used in the present investigation for Au+Au collisions at $\sqrt{s_{NN}}$ = 200, 39, 27, and 19.6~GeV. 
The used AMPT model has the string-melting mechanism, and hadronic cascade both turned on.  The AMPT model has been widely employed to investigate relativistic heavy-ion collision physics~\cite{Lin:2004en,Ma:2016fve,Ma:2013gga,Ma:2013uqa,Bzdak:2014dia,Nie:2018xog,Bzdak:2014dia,Magdy:2020xqs,Magdy:2022ize,Magdy:2022jai,Magdy:2022cvt}.
In the AMPT model with string melting on, the HIJING model is used for hadrons creations. These hadrons are then transformed into their valence quarks and anti-quarks. In addition, their time and space evolution is evaluated by the ZPC Parton cascade model~\cite{Zhang:1997ej}.

The AMPT has four essential components, (i) HIJING model~\cite{Wang:1991hta,Gyulassy:1994ew} for the initial Parton-production stage, (ii) the parton-scattering stage, (iii) hadronization through coalescence followed by (iv)  a hadronic interaction stage~\cite{Li:1995pra}.  
In the stage of parton-scattering, the parton-scattering cross section is given as;
\begin{eqnarray} \label{eq:21}
\sigma_{pp} &=& \dfrac{9 \pi \alpha^{2}_{s}}{2 \mu^{2}},
\end{eqnarray}
where $\mu$=$4.6$ gives the partonic matter screening mass and $\alpha_{s}$=$0.47$ represents the QCD coupling constant. The parameters $\mu$ and $\alpha_{s}$ generally give the expansion dynamics of the A--A collision systems~\cite{Zhang:1997ej,Xu:2011fi,Nasim:2016rfv,Solanki:2012ne}. In the current work, the $\sigma_{pp}$ is fixed to $1.5$~$mb$. 

In the current work, the centrality intervals are given by cutting on the charged particle multiplicity in midrapidity. Then the AMPT simulated events are analyzed using the multi-particle cumulant method~\cite{Bilandzic:2010jr,Bilandzic:2013kga,Jia:2017hbm,Gajdosova:2017fsc} using particles with pseudorapidities $|\eta| < 1$ and with  transverse momentum $0.2 < p_T < 2.0$~GeV/$\textit{c}$. 

The multi-particle cumulant technique is used for my correlation analysis. The framework of the multi-particle cumulant using one and many sub-events is described in Refs~\cite{Bilandzic:2010jr,Bilandzic:2013kga,Gajdosova:2017fsc,Jia:2017hbm}. I used the two-, three-, and four-particle correlations in this work using the two sub-events cumulant technique~\cite{Jia:2017hbm}. The two sub-events $\textit{A}$ and $\textit{B}$ with $|\Delta\eta|~ > 0.7$ (\textit{i.e.}, $\eta_{A}~ > 0.35$ and $\eta_{B}~ < -0.35$).  Using the two sub-events method helps reduce the non-flow correlations~\cite{Magdy:2020bhd}. The two-, three-, and four-particle correlations are given using the two sub-events cumulant method~\cite{Jia:2017hbm} as:
\begin{eqnarray}\label{eq:2-1}
v^{\rm Inclusive}_{n}                        &=&  v_{n} = \langle  \langle \cos (n (\varphi^{A}_{1} -  \varphi^{B}_{2} )) \rangle \rangle^{1/2},
\end{eqnarray}
\begin{eqnarray}\label{eq:2-2}
C_{n+m,nm}                                   &=&   \langle \langle \cos ( (n+m) \varphi_{1}^{A} - n \varphi_{2}^{B} -  m \varphi_{3}^{B}) \rangle \rangle ,
\end{eqnarray}
\begin{eqnarray}\label{eq:2-3}
\langle v_{n}^{2} v_{m}^{2}  \rangle &=& \langle \langle \cos ( n \varphi^{A}_{1} + m \varphi^{A}_{2} -  n \varphi^{B}_{3} -  m \varphi^{B}_{4}) \rangle \rangle,
\end{eqnarray}
where $\langle \langle \, \rangle \rangle$ represents the average over all particles and all events and $\varphi_{i}$ is the $\textit{i}^{th}$ particle azimuthal angle.

Using  Eqs.~(\ref{eq:2-1})-(\ref{eq:2-3}), the  mode-coupled response to the $\textit{v}_{n+m}$ can be given as~\cite{Yan:2015jma,Bhalerao:2013ina}:
\begin{eqnarray}\label{eq:2-4}
\textit{v}_{n+m}^{\rm MC} & =   & \frac{C_{n+m,nm}} {\sqrt{\langle \textit{v}_n^2 \textit{v}_m^2 \rangle}}, \\ 
                          &\sim & \langle \textit{v}_{n+m} \, \cos ((n+m) \Psi_{n+m} - n\Psi_{n} - m\Psi_{m}) \rangle.  \nonumber \\
\end{eqnarray}
Moreover, the linear response to $\textit{v}_{n+m}$ is given as:
\begin{eqnarray}\label{eq:2-5}
v_{n+m}^{\rm Linear}  &=& \sqrt{ (v^{\rm Inclusive}_{n+m})^{\,2} - (v^{\rm MC}_{n+m})^{\,2}  }.
\end{eqnarray}

The ratio of the mode-coupled response to the inclusive $v_{n+m}$ gives the correlations between different order flow symmetry planes;
\begin{eqnarray}\label{eq:2-6}
\rho_{n+m,nm} &=& \frac{v^{\rm MC}_{n+m}}{v^{\rm Inclusive}_{n+m}} , \\  \nonumber
              &\sim & \langle  \cos ((n+m) \Psi_{n+m} - n \Psi_{n} - m \Psi_{m}) \rangle.
\end{eqnarray}

The mode-coupled response coefficient, which gives the coupling to the higher-order anisotropic flow harmonics and can be given as:
\begin{eqnarray}\label{eq:2-7}
\chi_{n+m,nm} &=& \frac{v^{\rm MC}_{n+m}} {\sqrt{\langle  v_{n}^{2} \, v_{m}^{2} \rangle}}.
\end{eqnarray}

\section{Results and discussion}\label{Sec:3}
\begin{figure}[!ht]
\vskip -0.16cm
\centering{
\includegraphics[width=0.99 \linewidth, angle=0]{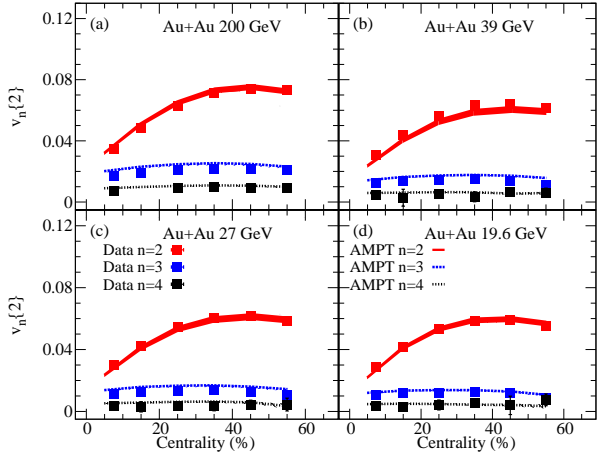}
\vskip -0.16cm
\caption{
Comparison of the experimental and simulated centrality and beam energy dependence of the $v_{n}\lbrace 2\rbrace$ for Au+Au collisions at 200~GeV panel (a), 39~GeV panel (b), 27~GeV panel (c), and 19.6~GeV panel (d). The solid points represent the experimental data reported by the STAR collaboration~\cite{STAR:2004jwm,STAR:2016vqt}.
\label{fig:fig1}
 }
}
\end{figure}

\begin{figure}[!ht]
\centering{
\includegraphics[width=0.99 \linewidth, angle=0]{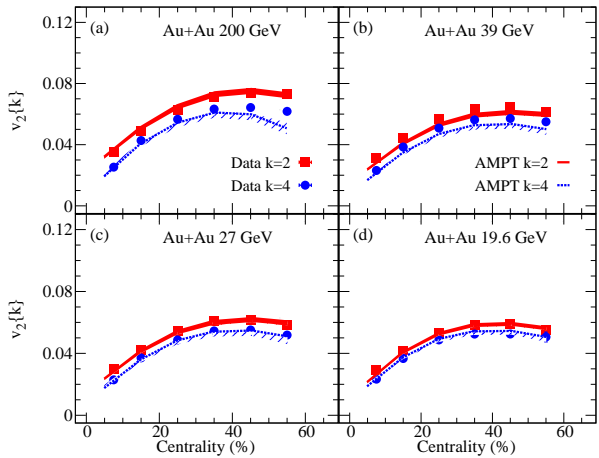}
\vskip -0.16cm
\caption{
The centrality-dependent of the $v_{2}\lbrace k\rbrace$ for Au+Au collisions at 200~GeV panel (a), 39~GeV panel (b), 27~GeV panel (c), and 19.6~GeV panel (d). The solid points represent the experimental data reported by the STAR collaboration~\cite{STAR:2004jwm,STAR:2016vqt}.
\label{fig:fig2}
 }
}
\end{figure}

Extracting the linear and the mode-coupled (i.e., non-linear) contributions to $v_4$ depends on the two- and four-particle correlations. Therefore, it is instructive to investigate the model's potential to simulate the experimental measurements of the two- and four-particle flow harmonics~\cite{STAR:2004jwm,STAR:2016vqt}.
Figures.~\ref{fig:fig1} and \ref{fig:fig2} show a comparison of the centrality dependence of  $v_{n}\lbrace 2 \rbrace$ and $v_{2}\lbrace 4 \rbrace$  for Au--Au collisions at 200 (a), 39 (b), 27 (c), and 19.6 (d)~GeV from the AMPT model. The AMPT calculations exhibit sensitivity to beam energy change and harmonic order $n$. They also indicate similar patterns to the data reported by the STAR experiment~\cite{STAR:2004jwm,STAR:2016vqt} (solid points). The data model comparisons suggest that the AMPT model contains the proper ingredient to describe the experimental data.
\begin{figure}[!ht]
\centering{
\includegraphics[width=0.99 \linewidth, angle=0]{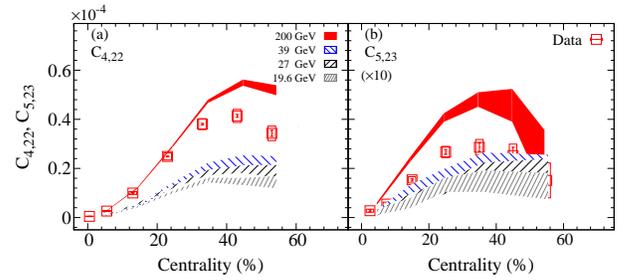}
\vskip -0.16cm
\caption{
The three-particle correlators, $C_{4,22}$ and $C_{5,23}$ centrality and beam energy dependence, from the AMPT model for Au+Au collisions. The points represent the experimental measurements at 200~GeV~\cite{Adam:2020ymj}.
\label{fig:fig3}
 }
}
\end{figure}
\begin{figure}[!ht]
\centering{
\includegraphics[width=0.99 \linewidth, angle=0]{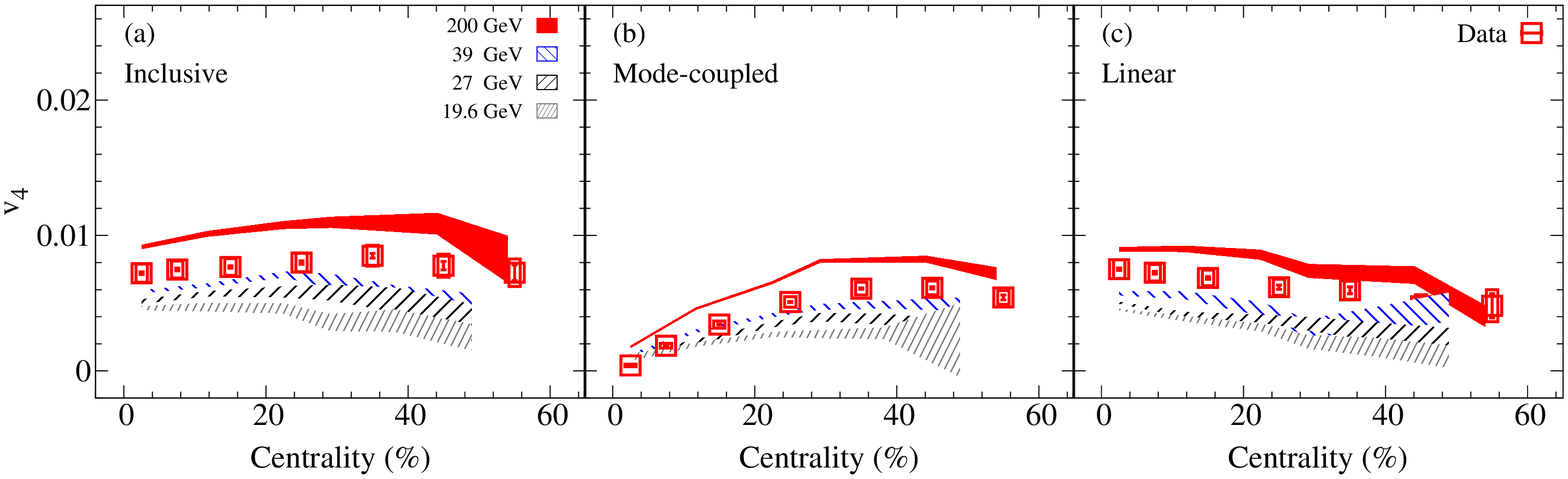}
\vskip -0.16cm
\caption{
Centrality and beam energy dependence of  the inclusive, non-linear and linear $\textit{v}_{4}$  obtained with the two sub-events cumulant method from the AMPT model for Au--Au collisions at 200~GeV. The points represent the experimental measurements at 200~GeV~\cite{Adam:2020ymj}.
\label{fig:fig4}
 }
}
\end{figure}
\begin{figure}[!ht]
\centering{
\includegraphics[width=0.99 \linewidth, angle=0]{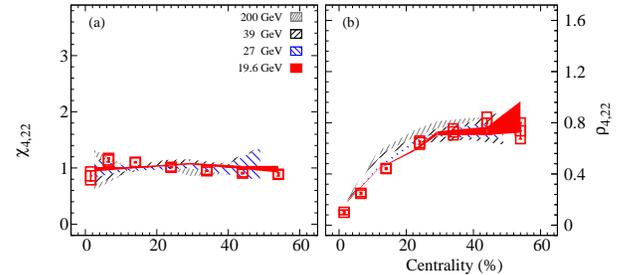}
\vskip -0.16cm
\caption{
Comparison of the $\chi_{4,22}$ and $\rho_{4,22}$ obtained from the AMPT model for Au--Au collisions at 200, 39, 27, and 19.6~GeV, as a function of centrality.The points represent the experimental measurements at 200~GeV~\cite{Adam:2020ymj}.
\label{fig:fig5}
 }
}
\end{figure}

The centrality dependence of the three-particle correlators, $C_{4,22}$ panel (a) and $C_{5,23}$ panel (b),  are shown in Fig.~\ref{fig:fig3} for Au+Au collisions at 200, 39, 27, and 19.6~GeV from the AMPT model. My results  demonstrate that $C_{4,22}$ and $C_{5,23}$ depend on beam energy. 
These dependencies indicate that $C_{4,22}$ and $C_{5,23}$ are susceptible to the change in viscous attenuation in AMPT (i.e., $\langle p_T \rangle$ and charged particle multiplicity) and the initial-state eccentricity. 
My results reflect the three-particle correlators' capability to constrain the interplay between the final- and initial-state effects in the AMPT model.
The AMPT  calculations qualitatively reproduce the trend observed in the experimental data~\cite{Adam:2020ymj}. However, AMPT overestimates $C_{4,22}$ for centrality larger than  30\% and $C_{5,23}$ for mid-central (10-50\%) region. 

Figure~\ref{fig:fig4} shows the centrality and beam energy dependence of the inclusive (a),  mode-coupled, and linear  $\textit{v}_{4}$ for Au+Au collisions from the AMPT model. My results indicated that the linear contribution is the dominant contribution to the inclusive $\textit{v}_{4}$ in central collisions at all presented energies. In addition, the $\textit{v}^{Linear}_{4}$ showed a weak centrality dependence. 
In addition, the difference between linear and mode-coupled $\textit{v}_{4}$ in central collisions is derived from the difference in $\varepsilon_{4}$ and $\varepsilon^{2}$, respectively.  
The presented results show that the inclusive, linear, and mode-coupled $\textit{v}_{4}$ are sensitive to the beam energy variation. 
The AMPT results are in qualitative agreement with the experimental measurements from the STAR experiment in Au+Au collisions at $\sqrt{s_NN}$ = 200~GeV~\cite{Adam:2020ymj,STAR:2022vqw}.

The mode-coupling response coefficient $\chi_{4,22}$, which gives the coupling strength between the lower and higher flow harmonics, is presented in Fig.~\ref{fig:fig5} panel (a) as a function of centrality for Au+Au at 200, 39, 27, and 19.6 GeV.  The $\chi_{4,22}$ calculations indicate a weak centrality and beam energy dependence,  which implies that (i) the $\chi_{4,22}$ is dominated by the initial-state eccentricity couplings and (ii) the mode-couped $\textit{v}_{4}$ centrality and energy dependence arise from the lower-order flow harmonics.
Figure~\ref{fig:fig5} panel (b) illustrates the centrality and energy dependence of the correlation between flow symmetry planes, $\rho_{4,22}$, in Au+Au collisions at $\sqrt{\textit{s}_{NN}}$ = 200, 39, 27, 19.6~GeV from the AMPT model.
The AMPT calculations of the $\rho_{4,22}$ indicate stronger event plane correlations in peripheral collisions for all presented energies. Nevertheless, the $\rho_{4,22}$ magnitudes are shown to be independent of beam energies. 
Such observation implies that initial-state eccentricity direction correlations dominate the correlation between flow symmetry planes.
In addition, these calculations are in agreement with the STAR experiment measurements in Au+Au collisions at $\sqrt{s_NN}$ =  200~GeV~\cite{Adam:2020ymj,STAR:2022vqw}. 


\section{Conclusion}\label{Sec:4}
I have presented comprehensive AMPT model calculations to evaluate the beam energy dependence of the linear and mode-coupling contributions to the $\textit{v}_{4}$, $\chi_{4,22}$ and $\rho_{4,22}$. 
The AMPT calculations indicate similar patterns and values to the experimental measurements of the $v_{n}\lbrace 2 \rbrace$ and $v_{2}\lbrace 4 \rbrace$.
The AMPT calculations of the mode-coupled $\textit{v}_{4}$ indicate a strong centrality dependence; however, it shows a weak centrality dependence for the linear $\textit{v}_{4}$.
In addition, the three particle correlations and the $\textit{v}_{4}$ show a strong beam energy dependence. In contrast, $\chi_{4,22}$ and $\rho_{4,22}$ show magnitudes and trends that are weakly dependent on beam energy.
The AMPT model calculations suggest that initial-state effects might be the dominant factors behind the correlations of event plane angles and the non-linear response coefficients.
\section*{Acknowledgments}
This research is supported by the US Department of Energy, Office of Nuclear Physics (DOE NP), under contracts DE-FG02-87ER40331.A008.
\bibliography{ref} 
\end{document}